\documentclass[conference]{IEEEtran}
\usepackage{cite}
\usepackage{amsmath,amssymb,amsfonts}
\usepackage{algorithmic}
\usepackage{graphicx}
\usepackage{textcomp}
\usepackage{xcolor}
\def\BibTeX{{\rm B\kern-.05em{\sc i\kern-.025em b}\kern-.08em
    T\kern-.1667em\lower.7ex\hbox{E}\kern-.125emX}}

\usepackage{subcaption}
\usepackage{todonotes}

\usepackage{url}

\ifx \showCODEN    \undefined \def \showCODEN     #1{\unskip}     \fi
\ifx \showDOI      \undefined \def \showDOI       #1{#1}\fi
\ifx \showISBNx    \undefined \def \showISBNx     #1{\unskip}     \fi
\ifx \showISBNxiii \undefined \def \showISBNxiii  #1{\unskip}     \fi
\ifx \showISSN     \undefined \def \showISSN      #1{\unskip}     \fi
\ifx \showLCCN     \undefined \def \showLCCN      #1{\unskip}     \fi
\ifx \shownote     \undefined \def \shownote      #1{#1}          \fi
\ifx \showarticletitle \undefined \def \showarticletitle #1{#1}   \fi
\ifx \showURL      \undefined \def \showURL       {\relax}        \fi
\providecommand\bibfield[2]{#2}
\providecommand\bibinfo[2]{#2}
\providecommand\natexlab[1]{#1}
\providecommand\showeprint[2][]{arXiv:#2}

\begin{document}

\title{Image-based Social Sensing: Combining AI and the Crowd to Mine Policy-Adherence Indicators from Twitter}

\author{\IEEEauthorblockN{Virginia Negri\IEEEauthorrefmark{1}, Dario Scuratti\IEEEauthorrefmark{1},  Stefano Agresti\IEEEauthorrefmark{1}, Donya Rooein\IEEEauthorrefmark{1}, Gabriele Scalia\IEEEauthorrefmark{1},
\\
Amudha Ravi Shankar\IEEEauthorrefmark{2}, Jose Luis Fernandez Marquez\IEEEauthorrefmark{2}, \\
Mark James Carman\IEEEauthorrefmark{1} and Barbara Pernici\IEEEauthorrefmark{1} (corresponding author)}
\IEEEauthorblockA{\IEEEauthorrefmark{1}Politecnico di Milano, Milan, Italy}
\IEEEauthorblockA{\IEEEauthorrefmark{2}Citizen Cyberlab, University of Geneva, Geneva, Switzerland}
}


\maketitle

\begin{abstract}
Social Media provides a trove of information that, if aggregated and analysed appropriately can provide important statistical indicators to policy makers. In some situations these indicators are not available through other mechanisms. For example, given the ongoing COVID-19 outbreak, it is essential for governments to have access to reliable data on policy-adherence with regards to mask wearing, social distancing, and other hard-to-measure quantities. In this paper we investigate whether it is possible to obtain such data by aggregating information from images posted to social media. The paper presents VisualCit, a pipeline for image-based social sensing combining recent advances in image recognition technology with geocoding and crowdsourcing techniques. Our aim is to discover in which countries, and to what extent, people are following COVID-19 related policy directives.
We compared the results with the indicators produced within the CovidDataHub behavior tracker initiative. Preliminary results shows that social media images can produce reliable indicators for policy makers.

\end{abstract}


\section{Introduction}
\label{sec:introduction}

The massive number of images posted to social media each day\footnote{Five hundred million tweets are posted to Twitter each day in 2020 according to: \url{https://www.webfx.com/internet-real-time/}
} 
represents a relatively untapped resource for mining useful social indicators and policy information. Providing better and more timely information to policy makers could provide widespread benefit by allowing for more reliable evidence-based decision making~\cite{fritz2019citizen}.

While text in social media has been mined extensively in the past, images have seen less interest, likely due to the difficulty to extract semantic information from them. Indeed according to a recent survey on social sensing~\cite{wang2019age}, one current research challenge is that of analysing the interdependent relationship between sensing measurements with different data modalities, such as text, sound, images, and
video, in order to obtain more accurate sensing results. In our paper we focus on jointly analysing text and images from social media posts.

Recent advances in deep-learning based image-processing techniques mean that the salient information contained within images is becoming easier to extract~\cite{DBLP:journals/nature/LeCunBH15}. Moreover, since each picture can contain a wealth of information (particularly compared to the limited amount of text that often accompanies it in platforms such as Twitter), we believe that image-based pipelines could substantially increase the breadth and depth of questions that can be answered using social media data.

The goal of this paper is thus to propose the  \emph{VisualCit (Visual Citizen) methodology for machine-learning enabled image-based social sensing of social indicators}. The methodology makes use of 
information from social media and processes it with 
\emph{AI and crowdsourcing} to discover and validate observations of social behaviour, which are then aggregated in now-casting fashion~\cite{2013:nowcasting} to estimate \emph{statistical indicators of social behaviour} that are useful to policy makers.
In our work, a semi-automated social sensing pipeline is developed that combines automated image classification techniques with crowd based validation techniques, which then allows for reliable estimation of social indicators. 

\begin{figure}[t]
    \centering
    \includegraphics[width=.3\columnwidth, height=.2\columnwidth]{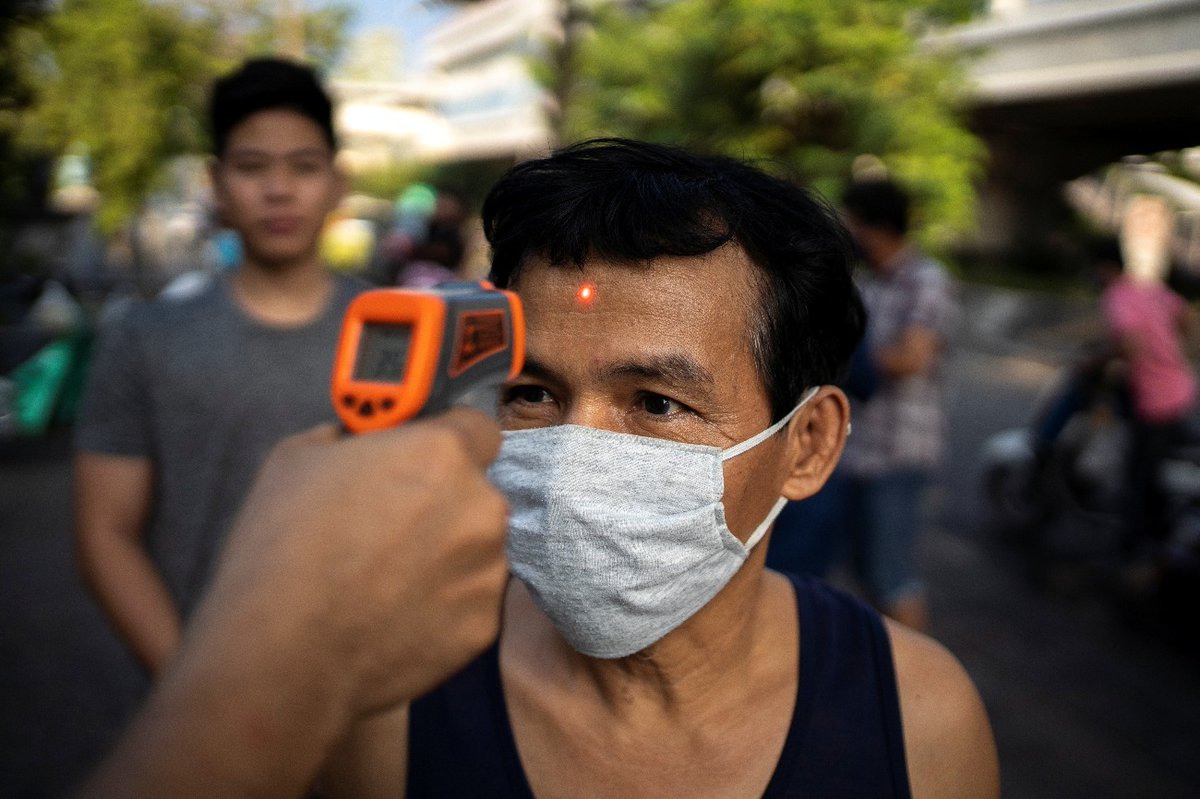}
    \includegraphics[width=.3\columnwidth, height=.2\columnwidth]{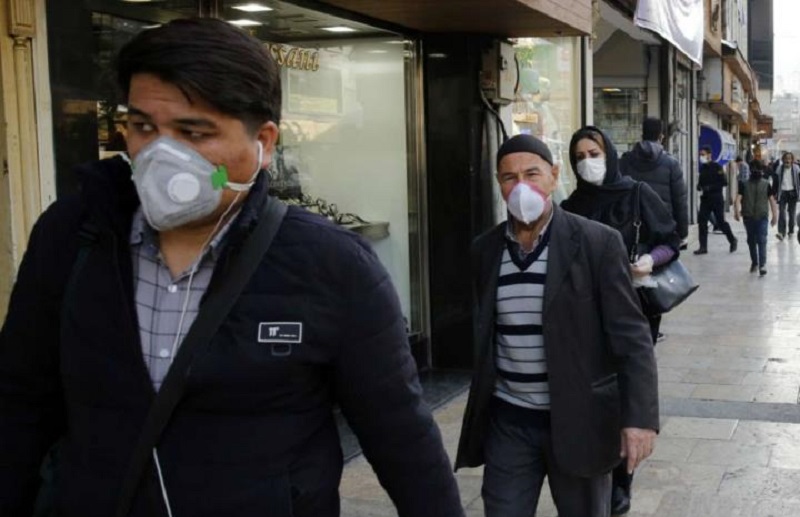}
    \includegraphics[width=.3\columnwidth, height=.2\columnwidth]{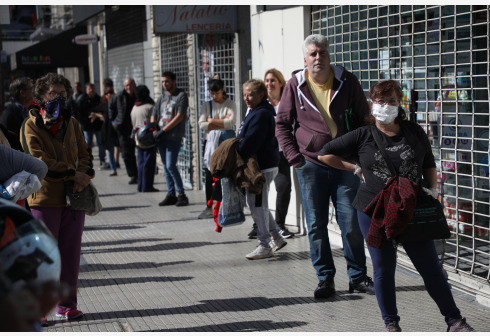}
    \includegraphics[width=.3\columnwidth, height=.2\columnwidth]{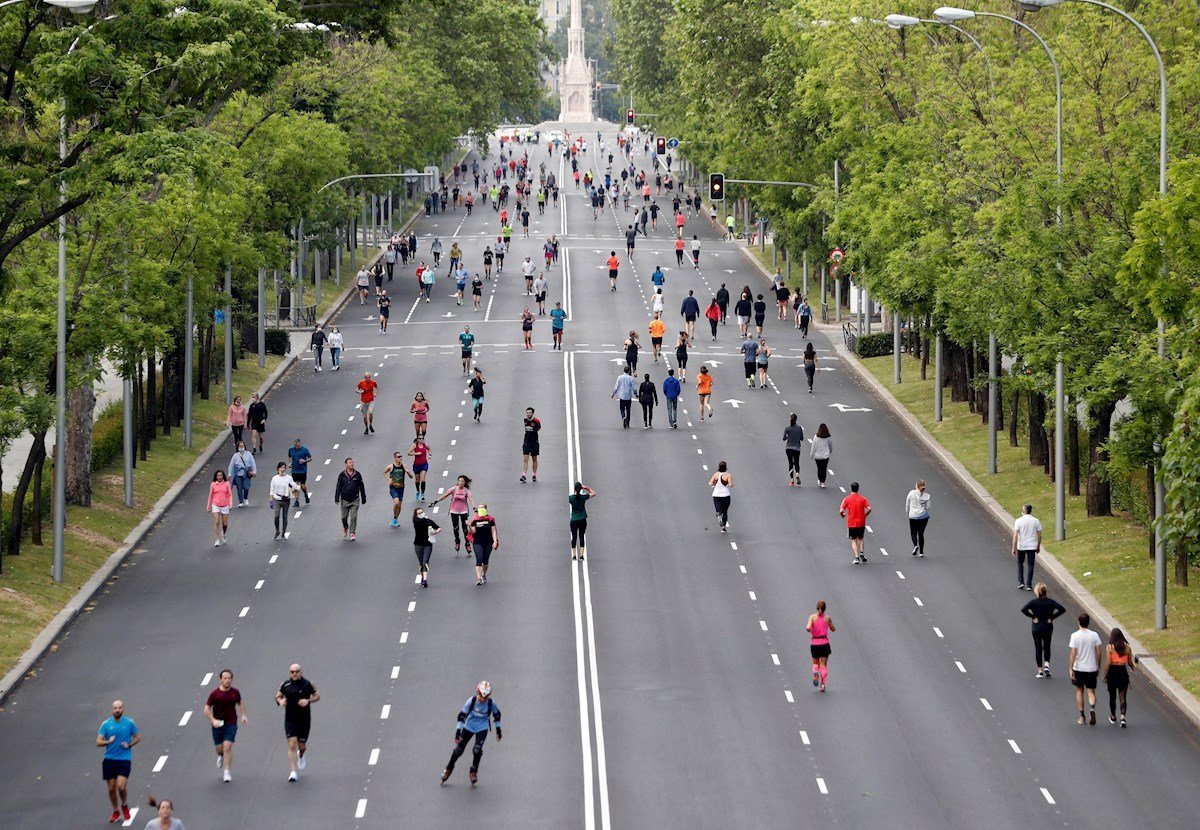}
    \includegraphics[width=.3\columnwidth, height=.2\columnwidth]{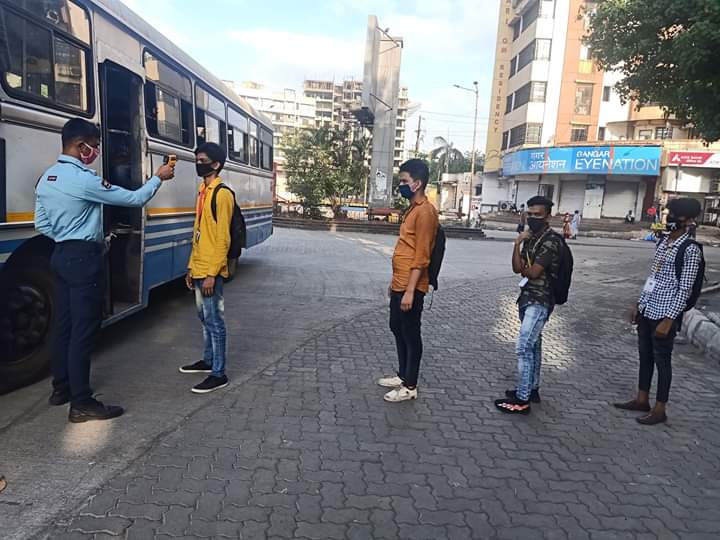}
    \includegraphics[width=.3\columnwidth, height=.2\columnwidth]{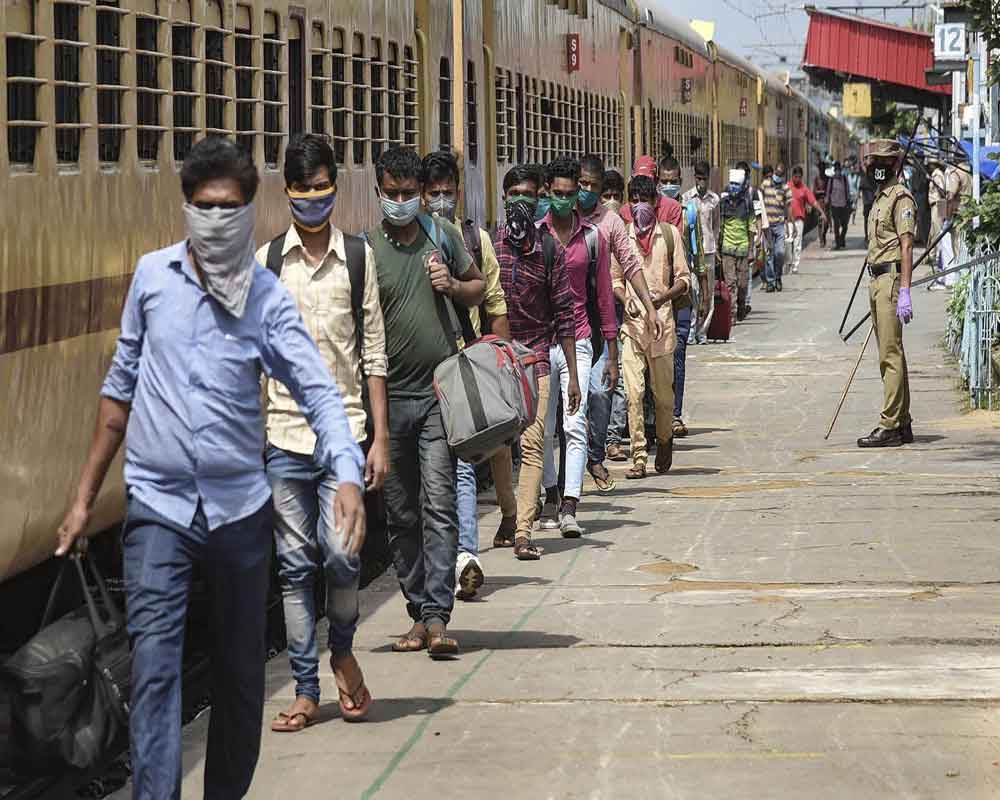}
    \includegraphics[width=.3\columnwidth, height=.2\columnwidth]{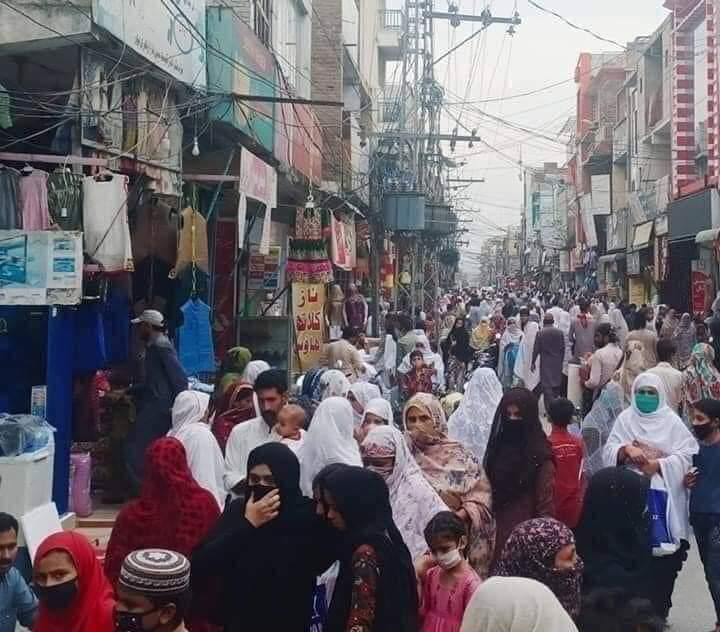}
    \includegraphics[width=.3\columnwidth, height=.2\columnwidth]{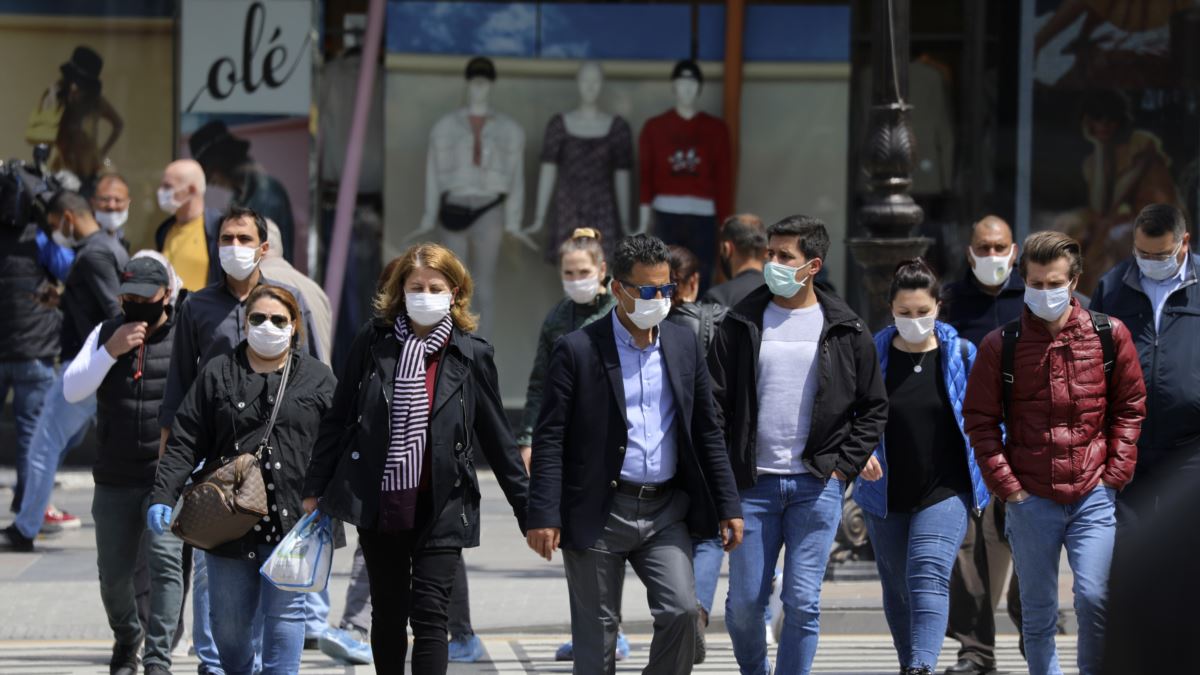}
    \includegraphics[width=.3\columnwidth, height=.2\columnwidth]{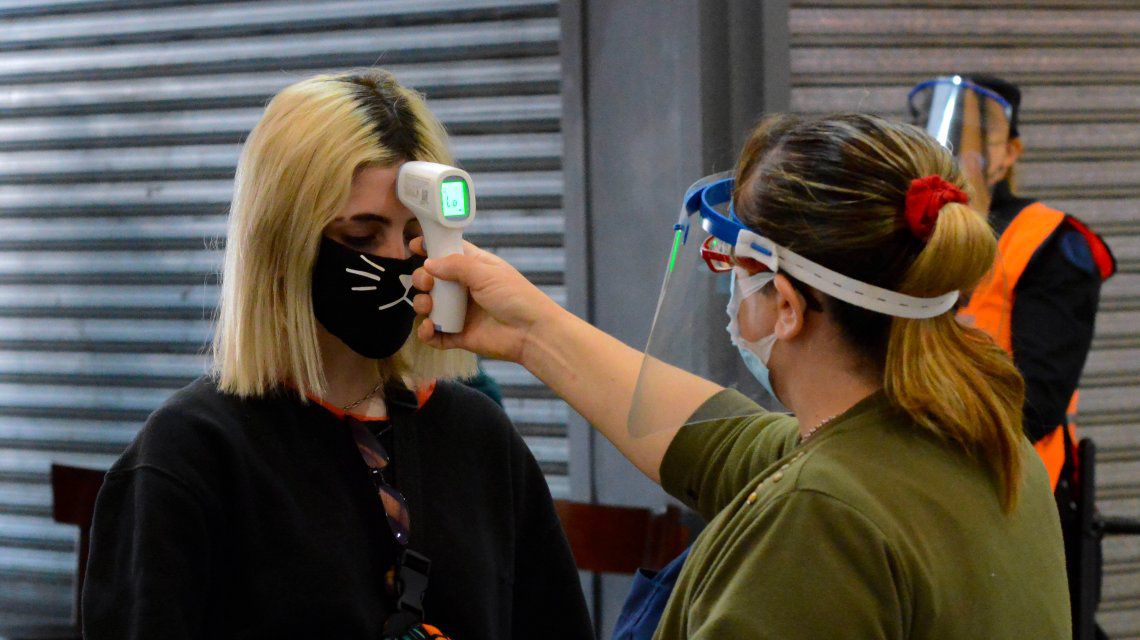}
    \caption{Examples of images posted to Twitter that could be useful for determining the level of COVID-19 related mask usage in different locations.}
    \label{fig:twitter}
\end{figure}

The specific application we tackle in this paper is that of monitoring indicators related to COVID-19 related policy directives such as the requirements to social distance and to wear masks.
Images on Twitter, such as those shown in Figure~\ref{fig:twitter}, provide useful information to an analyst tasked with the problem of determining the amount of policy-adherence in different locations. Obviously a single analyst cannot alone monitor the massive flow of images on Twitter to determine the amount of adherence, but by leveraging the image classification techniques we can automatically extract \emph{only} the relevant images from that massive stream. The resulting stream of relevant images may still be too large for a single analyst to deal with, but by recruiting a team of analysts through crowdsourcing, the capacity of the analysts can scale to fit the data need.   

In this work we combine the strengths of automated Machine Learning based image filtering techniques, namely its speed and scalability, with those of crowdsourcing, in particular its accuracy and flexibility, to present a new methodology for estimating policy indicators using an image-based social sensing framework to mine images from Twitter.

The main contributions of this work are as follows: 
\begin{itemize}
    \item We introduce a framework for \emph{image-based} social sensing that allows for image-based evidence of certain COVID-19 related social behaviour (mask wearing, social distancing, etc.) to be aggregated into indicators.
    \item We develop and test a series of image filters based on deep learning techniques to automatically select with high accuracy only those images which are photos depicting a certain type of event (e.g., 2 or more people meeting in a public place).
    \item We build a crowdsourcing application that leverages the crowd to extract the necessary information from the selected pictures for calculating the desired indicators.
    \item We derive indicators about COVID-related behaviors from the crowdsourcing results and compare them with data from other external sources, obtained through surveys \cite{coviddatahub}.
\end{itemize}

The paper is structured as follows. We first discuss related work in Section~\ref{sec:relatedwork}, presenting 
the state-of-the-art on 
several aspects relevant for developing an image-based social sensing pipeline. In Section~\ref{sec:approach} we present the VisualCit approach for extracting information from social media regarding the social impact of COVID-19. Specifically we develop a pipeline of tools to derive information from Twitter with an approach based on the  combination of AI and crowdsourcing. Finally, in Section \ref{sec:datasets} we illustrate an experimental dataset and evaluate the performance of step in the pipeline before comparing the resulting indicators with an external data source based on surveys.

\section{Related Work}
\label{sec:relatedwork} \label{sec:sota}

In this section we discuss the different branches of research that we bring together in this paper, namely social sensing, citizen science, and deep learning.

\subsection{Social Sensing versus Traditional Surveys}

\emph{Social sensing} has been proposed in the literature~\cite{2015:socialsensing} as a term
for describing the gathering of information from humans -- using crowdsourcing, human-connected devices (mobiles, etc.) and/or by extracting information from social media -- with the goal of mining social signals to gather situation awareness and support decision making.

The use of social media to gain timely evidence of ongoing emergency events (such of pictures of flooded towns, earthquake devastated buildings, or burnt forest, etc.) has been widely studied~\cite{havas2017e2mc}. Recent approaches propose to retrieve visual evidence on the events by combining both textual data mining and automated image classification, in order to reduce the information overload needed to inspect images manually~\cite{huang2019visual,IMRAN2020102261}. 

To the best of our knowledge, however, no previous work has looked to perform social sensing based on the visual information available in social media, i.e., to aggregate evidence from observations of social behaviour to compute near real-time social indicators that are useful for policy makers.

The traditional approach to collect policy adherence information is through surveys. In the context of the COVID-19 emergency, surveys regarding face mask usage have been performed\footnote{https://www.statista.com/statistics/1114375/wearing-a-face-mask-outside-in-european-countries/} and thematic maps have been produced based on survey results to study the evolution of mask use over time\footnote{\url{http://www.healthdata.org/sites/default/files/files/Projects/COVID/Mask_use_infographic_2020-1.pdf}}. 
The Covid-19 Behavior Tracker initiative (CovidDataHub project) by the Institute of Global Health Innovation (IGHI) at Imperial College London and YouGovSurveys collects data on COVID-19 behavioural aspects through surveys for a selection of countries. In order to generate the data, each week around 1,000 people from each country are interviewed, and summary data are made available for the countries in which the target number of respondents is reached.
While the total number of countries being surveyed is 30, 
the reports usually present data for a subset of countries, depending on the availability of data (e.g., only four countries were reported in the first week of August, 2020). These survey data are used as an external data source for validating the results of our pipeline. 

The limitation of survey-based social indicator estimation is, of course, the need to reach a \emph{sufficient number of representative individuals} on a regular basis. In this paper we propose a completely different approach based on social media and crowdsourcing that avoids altogether the need to find representative individuals and entice them to respond to online surveys. 

\subsection{Crowdsourcing and Citizen Science}

In 2014, the Oxford dictionary\footnote{https://www.oed.com/} integrated the term citizen science as: "scientific work undertaken by members of the general public, often in collaboration with or under the direction of professional scientists and scientific institutions". Different levels of participation and engagement have been defined from citizens as sensors to collaboration in project definition, data collection and analysis~\cite{haklay2013citizen}. 

The involvement of citizens in the solution of social science problems has been proposed and discussed in the literature. In the context of critical societal challenges, the authors of a recent roadmap paper~\cite{fritz2019citizen} discuss the difficulty of collecting data to measure the 232 indicators related to the Sustainable Development Goals (SDG) defined by the United Nations. Citizen-generated data are considered a possible non-traditional data source, that could be used to complement the official data sources which are often costly to produce in terms of both time and resources. The citizen-generated data often allows for wider coverage, both spatial and temporal. The collection process may different from traditional methods with the involvements of citizens at different expertise levels, actively or passively (through social media) collecting data to support scientists, or even fostering co-creation initiatives. The main issue in this case is the quality of the collected data. In addition to general data quality metrics, as the ones proposed by ISO 25012 for data quality in general and ISO 19157 for the quality of spatial data, the quality issue has been studied in depth in the context of crowdsourcing, with several strategies for ensuring data quality for this type of data~\cite{daniel2018quality,2020:jin_survey}. In this paper we make use of simple majority-vote based crowdsourcing quality control techniques, deferring the implementation of more sophisticated techniques to future work.

Another important issue which arises when citizens are involved in the collection and/or assessment of data to support scientific projects is the size of the data to be analysed. In this case the task is to present the citizens with a manageable amount of data to analyse, and to select only the data that are relevant for the problem being studied.  
As noted in the previous section, AI techniques such as the use of automated classifiers can be employed to reduce the amount of information provided to the citizen scientists. We follow this approach in the development of our pipeline.

We note that there are emerging examples in the health care domain of even more collaborative approaches between AI and crowdsourcing, targeted in particular at helping communities at risk, where health-care providers and experts involved through crowdsourcing are actively supported by AI techniques (e.g., \cite{ruiz2019snakebite,andrew2021SnakeChallenge}). 

\subsection{Deep Learning for Image Filtering}

A focus of our work is on automatically analysing the visual evidence emerging from social media \emph{before} involving crowd workers and experts. Thus in this section we provide a brief overview of technology developments in deep learning based image processing that allow for implementing the large-scale filtering of images needed in this project. Later in Section \ref{sec:approach} we will illustrate our specific approach and the specific types of filters we have developed. 

As noted above, deep learning and in particular deep Convolutional Neural Network (CNN) architectures have massively improved the state-of-the-art performance in image recognition tasks (such as image classification, object recognition, segmentation, etc.) over the last few years~\cite{DBLP:journals/nature/LeCunBH15}. Models like VGG~\cite{2015:vgg}, ResNet~\cite{resnet} and EfficientNet~\cite{2019:efficientnet} come pre-trained on the massive image datasets, such as ImageNet~\cite{2015:imagenet}, and can then be fine-tuned on specific classification tasks for excellent performance, provided sufficient training data are available. In this paper, we make use of several pre-trained models. Moreover, in order to leverage task-specific training on large external collections, we often make use of models that have been fine-tuned and extended (in terms of the network architecture) for certain image processing tasks, for which we require a particular filter. In all cases, performance of these models could be further improved by trained on task specific data gathered during the crowdsourcing phase of our pipeline. 

The functionality required for building an image filtering pipeline for social sensing can be summarised in two key questions: 
\begin{enumerate}
    \item What type of objects does the image contain?
    \item Where was the photo taken?
\end{enumerate}

To answer the first question, it is important not only to look at which objects are displayed, but most importantly, whether the displayed content is safe to show to crowdworkers. For this purpose, specialised Not Safe For Work (NSFW) classifiers exist, such as Yahoo's OpenNSFW\footnote{\url{https://github.com/yahoo/open_nsfw}}. This model uses a convolutional neural network based on ResNet-50~\cite{resnet}.  

To identify specific objects in images, a number of techniques exist, with the most famous among them being YOLO \cite{yolo}, a real-time object detector pre-trained on the COCO \cite{coco} dataset. YOLO is fast and accurate, outperforming Faster-RNN~\cite{faster_rnn}, the previous state-of-the-art, both in terms of accuracy and speed (100 times faster). YOLO is a convolutional model that with a single pass is able to simultaneously predict multiple bounding boxes and class probabilities for each box. Thus, it can be trained end-to-end, differently from traditional region proposal networks, and as such is much faster and more accurate. 
The COCO dataset, on which YOLO is trained, provides a large number of object categories, allowing a filter based on it to be used in a wide variety of scenarios.
Since our study focuses on COVID-related social behaviour, an object detector can be used to filter out images containing less than two people. 

The second criterion for filtering the image content is to look at where the depicted event occurred. Scene classifiers can be used for this purpose, since we can select the types of locations of our interest.
This approach allows us to be flexible in the choice of the scene and select the ones pertinent to the goal of the study. An open source repository containing various convolutional neural networks (CNNs) pre-trained on Places365 dataset~\cite{places365} is available\footnote{\url{https://github.com/AMANVerma28/Indoor-Outdoor-scene-classification}}. This dataset gathers images belonging to 365 scene categories, which are sufficiently specific to be used in a wide range of tasks, (including in our case to detect whether the location is public or private).

\subsection{Geolocating Observations}

In the context on many social sensing projects that make use of data from social media, one of the important issues is the ability to associate a location to the information been extracted.
Considering Twitter as the most common social network from which social media information is extracted, one problem is that only a small percentage of tweets are natively geolocated and all images are stripped of metadata for privacy issues.
As a result, many authors have studied the problem of geolocating tweets from the available information (e.g., \cite{DBLP:journals/tkde/ZhengHS18,DBLP:journals/tois/MiddletonKPK18}).

In this paper we adopt the CIME geolocation algorithm proposed in the E2mC project \cite{francalanci2017exploratory,francalanci2018talking}, which for non-geolocated tweets extracts a possible location from the text and metadata of the post, using the Stanford Core Named Entity Extraction algorithm~\cite{DBLP:conf/acl/ManningSBFBM14} and OpenStreetMap~\cite{DBLP:journals/pervasive/HaklayW08} with the Nominatim API\footnote{\url{https://wiki.openstreetmap.org/wiki/Nominatim}} as a gazeteer and a context-based approach for  disambiguation \cite{francalanci2018talking}.

\subsection{Preventing Duplicate Observations}

It is important in our framework to have filters than can automatically detect and remove identical or similar images, since they would increase the workload for the crowd, and more importantly could bias and distort the estimates of social indicators through the double counting of individual observations. 

Detecting similar images that come from the same original photo source is non-trivial, since intermediate processing might in different spatial resolutions or the enhancement with various image filters or event cropping of the source image. It is possible to train Deep Convolutional Neural Networks in a so-called Siamese Architecture for detecting near duplicates~\cite{2005:similarity}. Simpler techniques that require no training, but instead make use of similarity preserving hash functions on images are also available, such as the well-known Perceptual Hash (P-hash) functions~\cite{2010:perceptualhashing}. Modifications to an image, such as the rescaling and enhancement mentioned above, can easily be detected by comparing the images' P-hash with previously seen values.

\begin{figure}[t]
    \centering
    \includegraphics[width=.35\columnwidth]{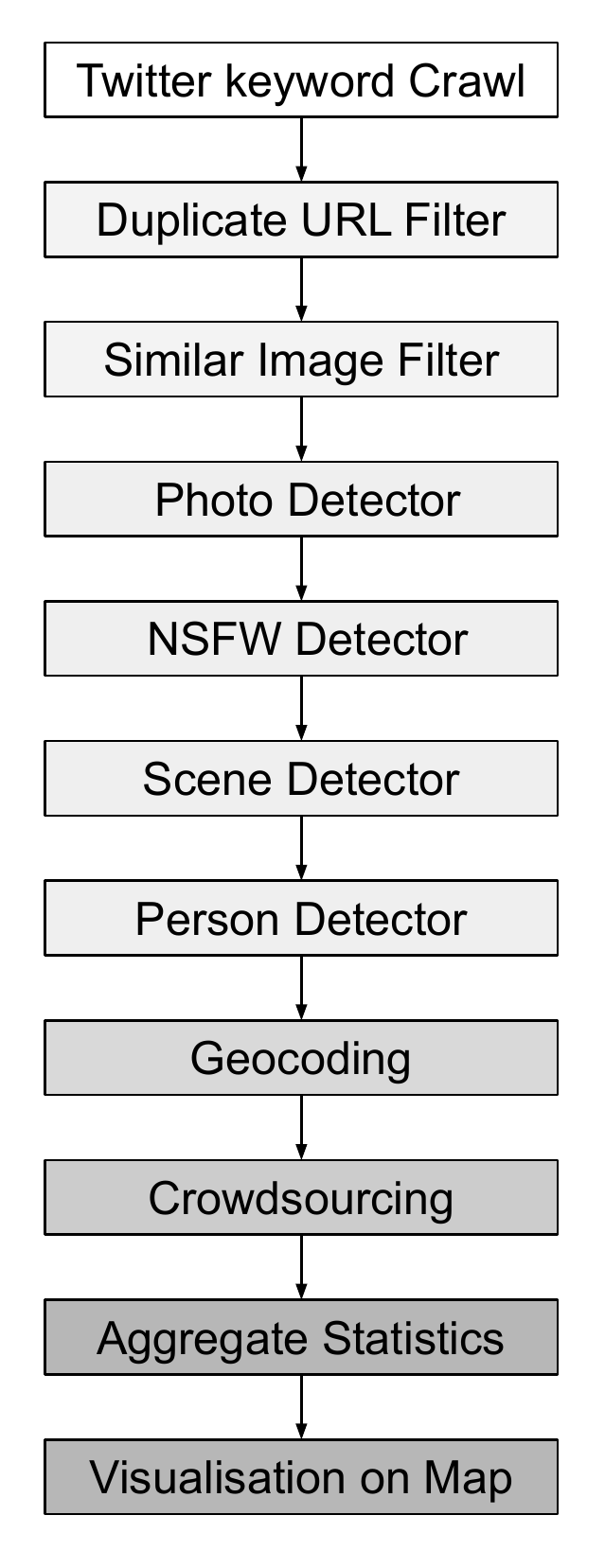}
    \vskip -.5em
    \caption{Components of the VisualCit social sensing pipeline.}
    \label{fig:pipeline}
\end{figure}

\section{The VisualCit Approach}
\label{sec:approach}

The VisualCit approach to image-based social sensing from Twitter leverages the respective strengths of machine learning-based automated image classification techniques and human user-based crowdsourcing. Figure~\ref{fig:pipeline} provides a depiction of the workflow developed. We now describe each of the components in the pipeline, focusing on the approach being followed to achieve the main goals: 
\begin{itemize}
\item
automatically extract and select images from Twitter posts that likely provide evidence for estimating a social indicator (e.g. images of people meeting in a public space),
\item
determine whether the candidate observations can be located  automatically (at the country level), and 
\item
ask the crowd to validate and annotate the candidate observations, such that they can be aggregated into estimates of the indicator. 
\end{itemize}
In the following subsections we illustrate the important steps in the pipeline. The code is available on request writing to the corresponding author.

\subsection{Keyword-based Crawling}

The Twitter crawls analysed in this paper (see Section \ref{sec:experiments}) were performed using the Tweepy library\footnote{https://www.tweepy.org/} to access the Twitter API\footnote{https://developer.twitter.com/en}. During the crawl only tweets containing images were retrieved and retweets were excluded. 


\subsection{Removing Duplicate Images}

Since each image represents a different observation for the purpose of estimating an indicator statistic (such as the amount of mask usage in a particular country), it is important that the same image does not pass multiple times through the pipeline. The same image of an event may be posted (or retweeted) by different people on social media, and thus may appear multiple times in our crawl. Thus at the very beginning of the pipeline we implement a check to remove duplicate image URLs from the crawl. 

\begin{figure}
\centering
\begin{subfigure}{0.5\columnwidth}
  \centering
  \includegraphics[width=0.9\columnwidth]{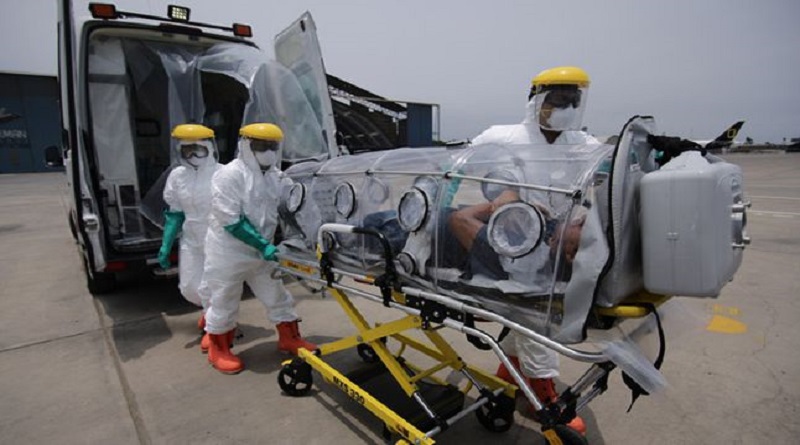}
  \caption{Original image}
  \label{fig:phash_0}
\end{subfigure}%
\begin{subfigure}{0.5\columnwidth}
  \centering
  \includegraphics[width=0.5\columnwidth]{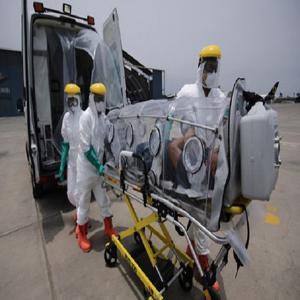}
  \caption{Rescaled image}
  \label{fig:phash_1}
\end{subfigure}
\begin{subfigure}{0.5\columnwidth}
  \centering
  \includegraphics[width=0.9\columnwidth]{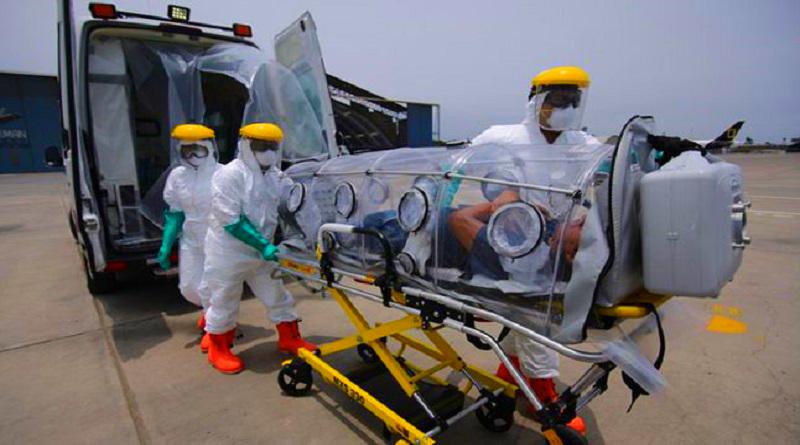}
  \caption{Saturated image}
  \label{fig:phash_2}
\end{subfigure}%
\begin{subfigure}{0.5\columnwidth}
  \centering
  \includegraphics[width=0.9\columnwidth]{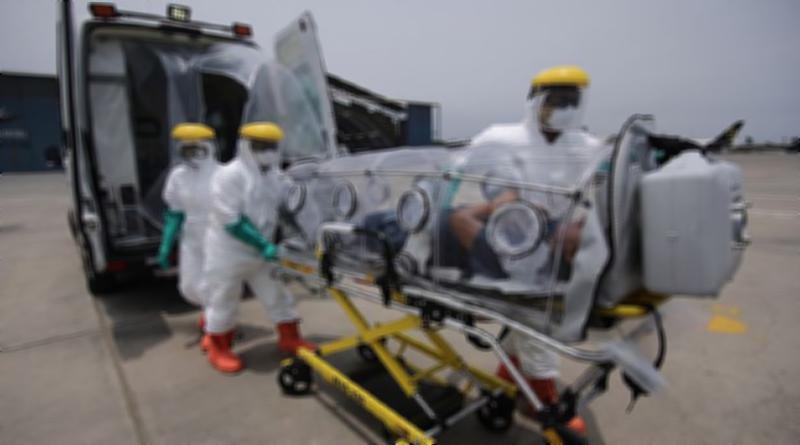}
  \caption{Blurred image}
  \label{fig:phash_3}
\end{subfigure}
\caption{Transformations on an image that do not affect its P-hash value 
}
\label{fig:phash_example}
\end{figure}

Checking for duplicate URLs does not guarantee the complete absence of duplicate images however, since the same image content could be available from different locations. One could remove duplicate images by computing their cryptographic hashes and discarding those already seen. However, this approach cannot detect if two images are  different but both come from the same original source image. The same image, for example, may have been uploaded with different spatial resolutions or been modified with colour filters. One way to allow for such transformations is to use a similarity-preserving hash function for comparing images. Similar images will have a same similarity hash and can thus be discarded.
In our framework we adopt the perceptual hash (P-hash)~\cite{2010:perceptualhashing} function. 
The P-hash algorithm extracts transformation-invariant features from the multimedia object and computes their hash. As a result, similar images will have the same features, and thus also the same hash, as shown for instance in Figure \ref{fig:phash_example}.

\subsection{Filtering Irrelevant Images}
\label{filtering_classifiers}
In order to fulfill our goal of gathering significant statistical information for policy makers, we must provide our crowd with only relevant data for the task. Thus we perform a set of filtering operations to extract only those images that are likely relevant. For this purpose, we built an image filtering pipeline based on deep learning techniques, including both state-of-the-art models pre-trained on large public datasets, and custom filters built according to our needs. The pipeline performs the following filter operations:
\begin{itemize}
    \item Removing non-photos
    \item Removing NSFW content
    \item Detecting the scene
    \item Detecting people
\end{itemize}
In principle, the four classifiers can be applied in any order, however, we ordered them on the basis of their performance characteristics, considering both speed of execution and selectivity. The characteristics of the image classification filters applied separately on 1,000 randomly selected images is reported in Table~\ref{tab:selectivity}. The geocoding algorithm requires more than a second per tweet, so it was performed after the image filtering.

\begin{table}[t]
\centering
\caption{Selectivity and execution time for image filters on 1,000 randomly chosen images.}
\begin{tabular}{lrr}
\textbf{Filter} & \textbf{images removed} & \textbf{time / image} \\ 
\hline
person detector & 78.6\% & 0.99s \\
photo detector & 65.3\% & 0.58s \\
NSFW detector & 7.7\% & 0.33s \\
public/private scene & 20.3\% & 0.34s \\
\hline
\end{tabular}
\vskip 1em
\label{tab:selectivity}
\end{table}

\subsubsection{Removing non-photos}
In order to efficiently remove from the crawled images those that do not represent photos, a photo detector was implemented. Crawled images contained a significant percentage of irrelevant images corresponding to internet memes or modified photos with text. To tackle this problem, a VGG19 model, pre-trained on the ImageNet dataset \cite{ImageNet}, was fine-tuned on a data set containing 3,376 images labelled as memes / non acceptable photos (taken from the Reddit Memes Dataset\footnote{\url{https://www.kaggle.com/sayangoswami/reddit-memes-dataset}}, and 2,448 images considered acceptable (taken from the Multi-Salient-Object (MSO) Dataset\footnote{\url{https://www.kaggle.com/jessicali9530/mso-dataset}}). To fine-tune the algorithm, VGG19's last layer was substituted to adapt the model for the new classification task, and all layers inherited from the original architecture are held frozen during training. In this way, the model achieved excellent performance on the filtering task. 

\subsubsection{Removing NSFW content}
It is critical to discard all \emph{Not Safe For Work} (NSFW) content from our data before feeding it to the crowd workers. To ensure this, we made use of Yahoo’s implementation of a NSFW classifier, OpenNSFW\footnote{\url{https://github.com/yahoo/open_nsfw}}.
Deciding what type of content is safe or not is subjective and context-specific. Yahoo's model specifically filters out pornographic content, while it does not address non-photos or offensive text, which we will both target by using a photo filter. It also does not address images depicting violence, which however we might want to include to investigate people's behaviour to help policy makers. We use this model as a preliminary filter, knowing that it provides a limited  guarantee on the accuracy of the output. We will thus necessarily warn our crowd on the probability of facing explicit content and ask to alert us in such a case. 

\subsubsection{Detecting the scene}
Selecting the right scene in images allows extracting a more meaningful subset of data for our task. For this purpose we introduced a scene detector in our pipeline.
This component consists in a convolutional neural network able to classify an image as belonging to one of a set of scene categories. In our framework we introduced an open-source model\footnote{https://github.com/AMANVerma28/Indoor-Outdoor-scene-classification} pre-trained on Places365 \cite{places365}, a public dataset of images corresponding to 365 scene categories.
For our specific task we thought a more meaningful distinction was between public and private scenes. Thus, we aggregated the original 365 scenes in these two subsets. Policy makers will surely be more interested in observing how people are behaving in public scenes such as streets or supermarkets, where they must conform to security regulations, rather than in private spaces.

\subsubsection{Detecting People}
We can greatly benefit from detecting required objects in a scene to narrow down the most relevant images for our purpose. For this purpose we introduced in our pipeline the YOLO \cite{yolo} object detector, pre-trained on the COCO dataset \cite{coco}. 
In the specific scenario of gathering relevant images for policy makers during the COVID-19 outbreak, we extracted images containing people. In addition, filtering images with at least two people showed a significant increase of the quality of the result, for example discarding  selfies.

\begin{figure}[t]
\centering
\begin{subfigure}{0.5\columnwidth}
  \centering
  \includegraphics[width=0.9\columnwidth]{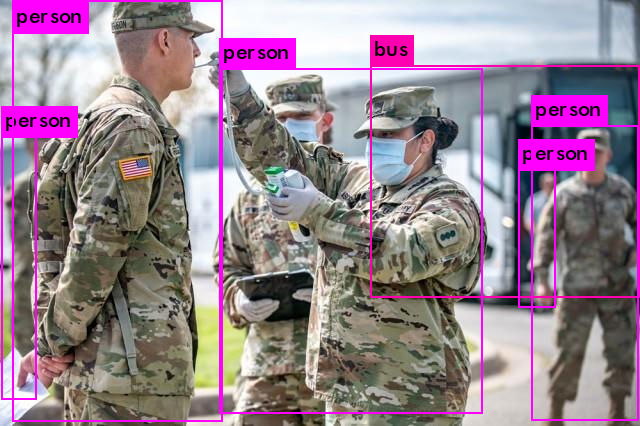}
  \label{fig:yolo_example_1}
\end{subfigure}%
\begin{subfigure}{0.5\columnwidth}
  \centering
  \includegraphics[width=0.9\columnwidth]{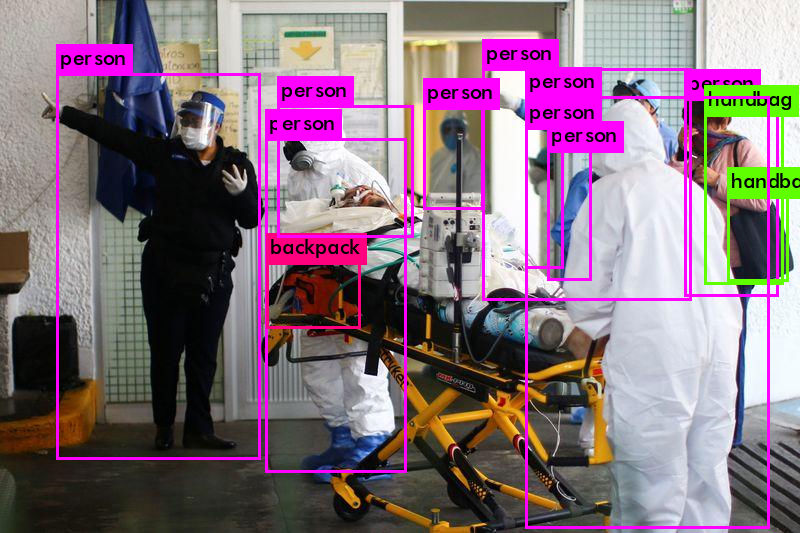}
  \label{fig:yolo_example_2}
\end{subfigure}
\caption{YOLO object detector used to detect people.}
\label{fig:yolo_example}
\end{figure}

\subsection{Geocoding Images}

In order to evaluate the social impact of COVID-19 in different countries, it is necessary to associate a location to each post.
The geolocation was performed using the CIME service \cite{francalanci2017exploratory,francalanci2018talking} described in Section \ref{sec:sota}, applying it on the textual part of the tweet, combined with the textual user location, if present. 
The geolocation was not performed whenever already available from Twitter itself, in which case the original one is used.

With the goal of creating thematic maps, we located each post with CIME and then extracted the country or territory it refers to. When multiple candidate locations were available, one has been chosen randomly to be shown to the crowd. The CIME function we used returns the coordinates of the centroid for each candidate location. We used this location to extract the corresponding country or territory code.  Only tweets with an associated location are sent to the crowd workers for the analysis, showing the name of the location as text.

\begin{figure}[th]
    \centering
    \includegraphics[width=1.0\columnwidth]{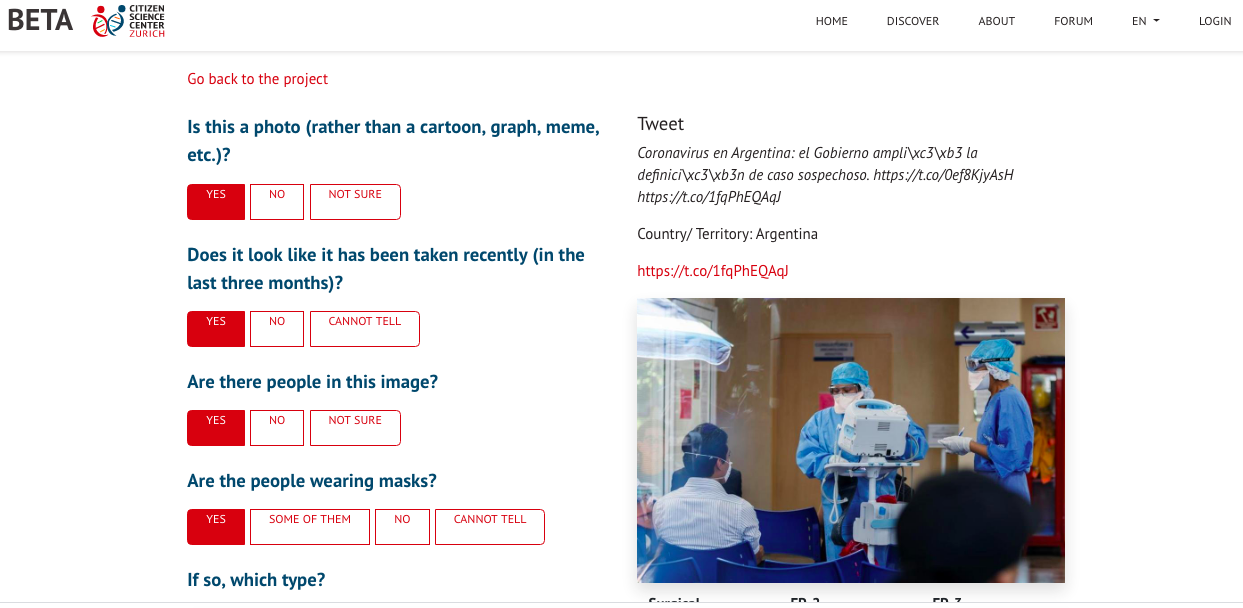}
    \caption{Crowdsourcing interface with the questions posed on the left and the text of the tweet, the proposed country and the image from the tweet on the right.}
    \label{fig:hospital}
\end{figure}

\subsection{Crowdsourcing}

In our project we use a citizen science approach to complement the collection of information asking the crowd to evaluate the behaviour of people in the extracted images. A series of questions is posed to the crowd workers, to assess the visible behaviour of the people. In particular we focused on social distancing and the use of face masks, as these data are difficult to extract automatically\footnote{See for example existing challenges on mask detection, e.g.  \url{https://www.aicrowd.com/challenges/mask-detection-challenge}}
and in many cases requires human judgement.

The open source PyBossa\footnote{\url{https://pybossa.com/}} platform for human data mining has been adopted in this project,
with the extension of the Project Builder realised at the Citizen Science Center Zurich for an easier creation and management of crowdsourcing projects. Each post is shown to the crowd worker with the image and a proposed geolocation for it (see Fig. \ref{fig:hospital}).
A series of questions concerning the image contents related to the Covid-19 pandemic are proposed to the crowd worker, concerning social distance and face mask usage. The full list of questions are listed in the Appendix~\ref{sec:appendix}. Some of the queries are conditioned on the previous question.\footnote{For instance, if the crowd worker's response to the first question: \emph{``Is this a photo (rather than a cartoon, graph, meme, etc.)?''} is \emph{``No''}, then no further questions are asked of that image.}

For each tweet, a separate task is created, and redundancy is set to 3 such that 
three independent crowd workers have to analyse the tweet for the task to complete. This is done in order to ensure and be able to assess the quality of the crowdsourcing results. The crowd was composed by 38 volunteers from social networks and working students.

\subsection{Result Aggregation and Visualisation}

For the completed tasks, the most frequent response is selected using a majority mechanism to proceed with the analysis. In addition all posts for which a \emph{``surely not''} answer was given for the question \emph{``Do you think the picture was likely taken in this location?''} were discarded as they are not useful for the mapping purpose as their geolocation is not correct.

After the crowd assessment and aggregation,  thematic maps are produced presenting the resulting indicators, i.e., percentages for the various questions across the countries for which sufficient data has be retrieved. 

\begin{figure}
\centering
\begin{subfigure}{0.5\columnwidth}
  \centering
  \includegraphics[width=0.9\columnwidth]{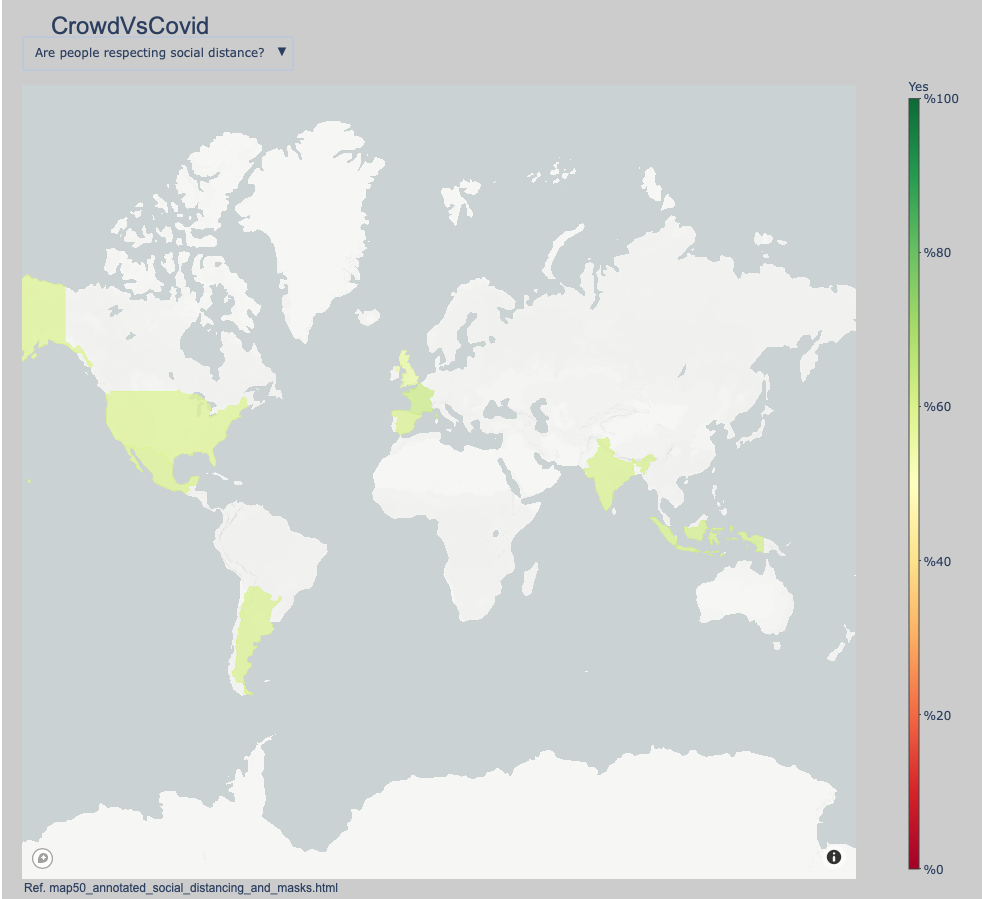}
\end{subfigure}%
\begin{subfigure}{0.5\columnwidth}
  \centering
  \includegraphics[width=0.9\columnwidth]{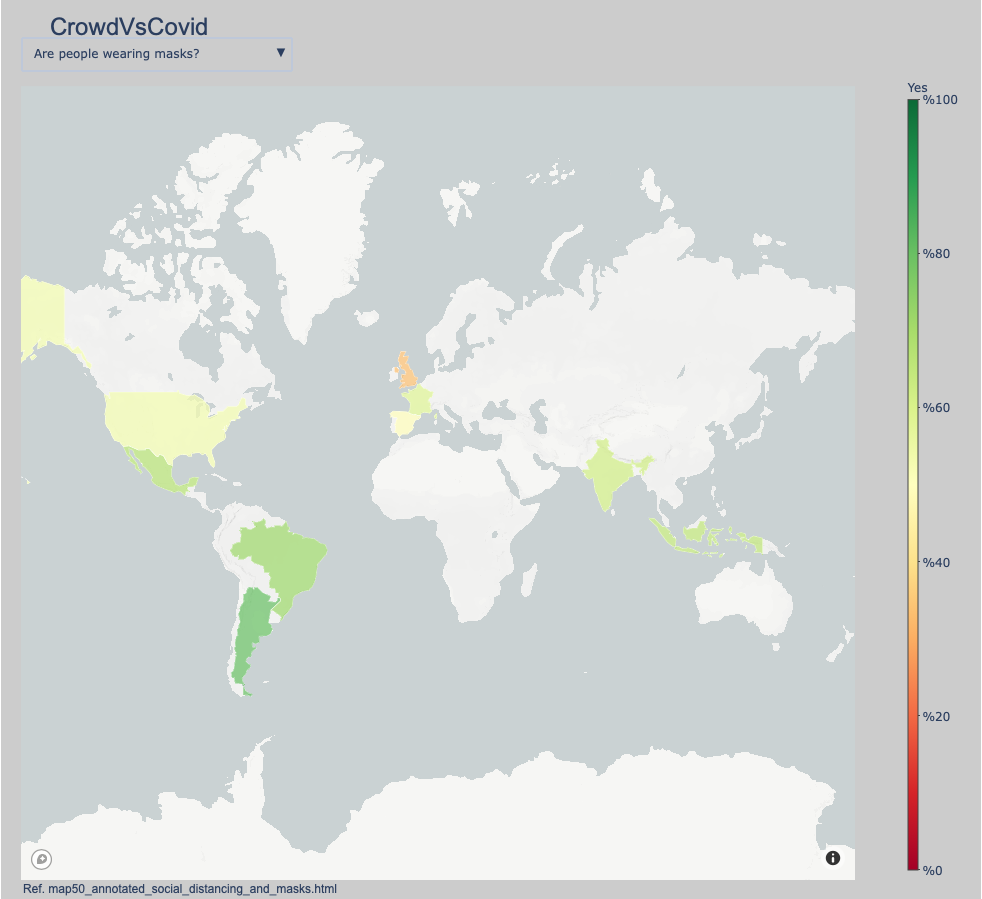}
\end{subfigure}
\begin{subfigure}{0.5\columnwidth}
  \centering
  \includegraphics[width=0.9\columnwidth]{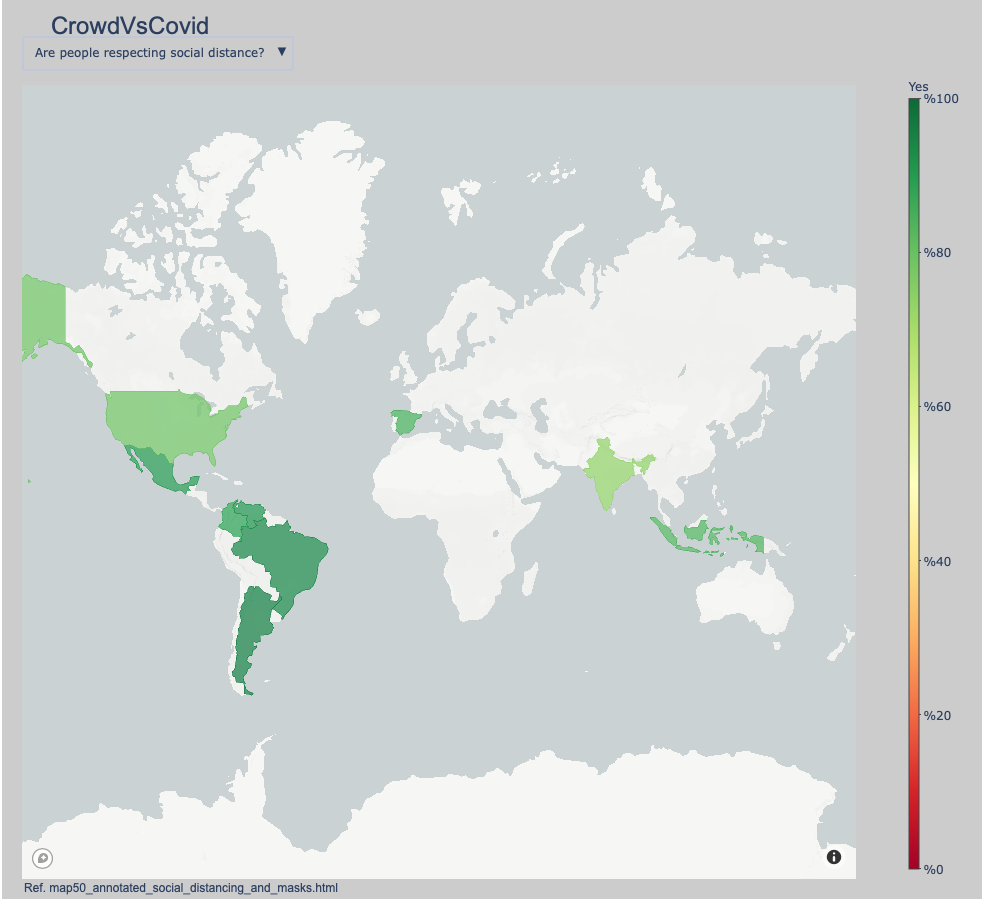}
\end{subfigure}%
\begin{subfigure}{0.5\columnwidth}
  \centering
  \includegraphics[width=0.9\columnwidth]{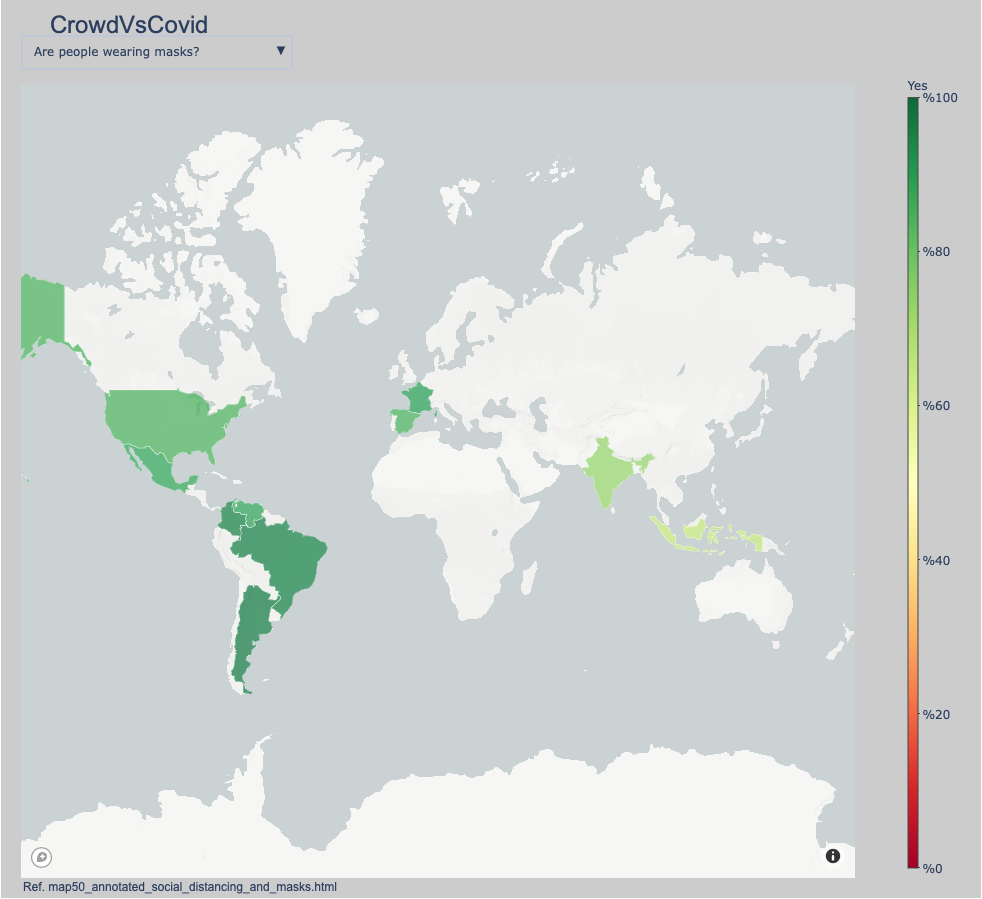}
\end{subfigure}
\begin{subfigure}{0.5\columnwidth}
  \centering
  \includegraphics[width=0.9\columnwidth]{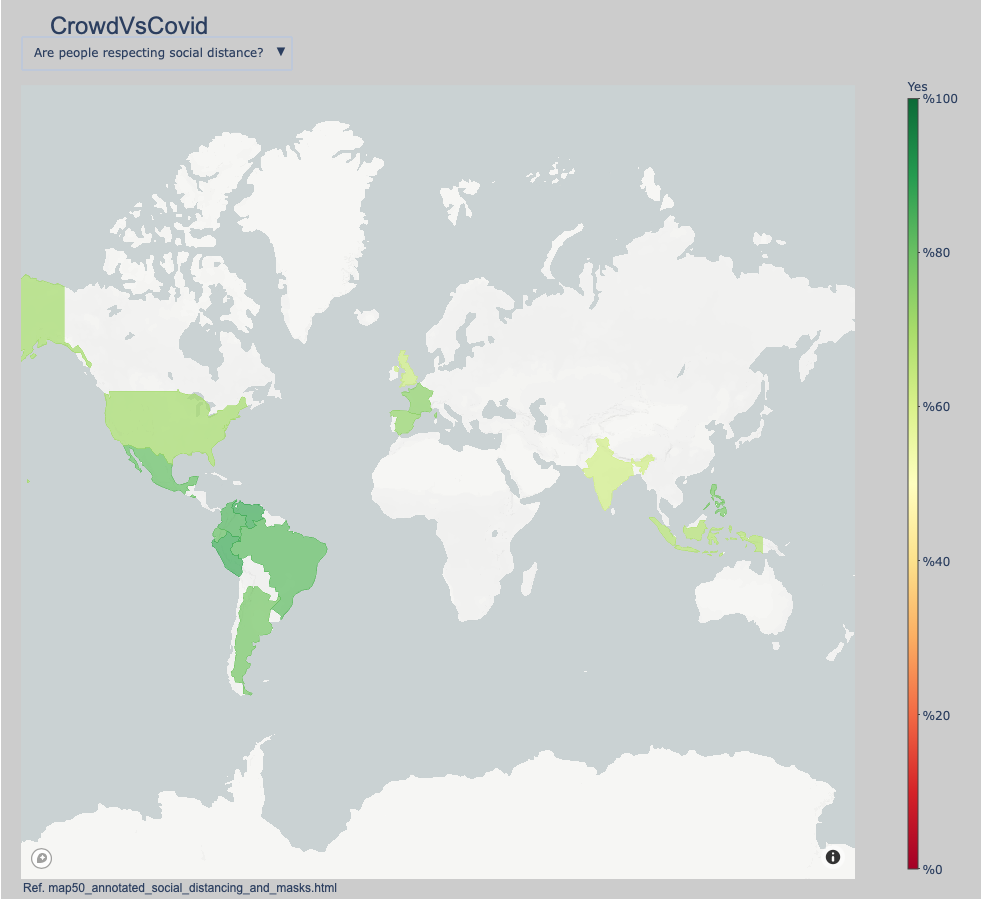}
\end{subfigure}%
\begin{subfigure}{0.5\columnwidth}
  \centering
  \includegraphics[width=0.9\columnwidth]{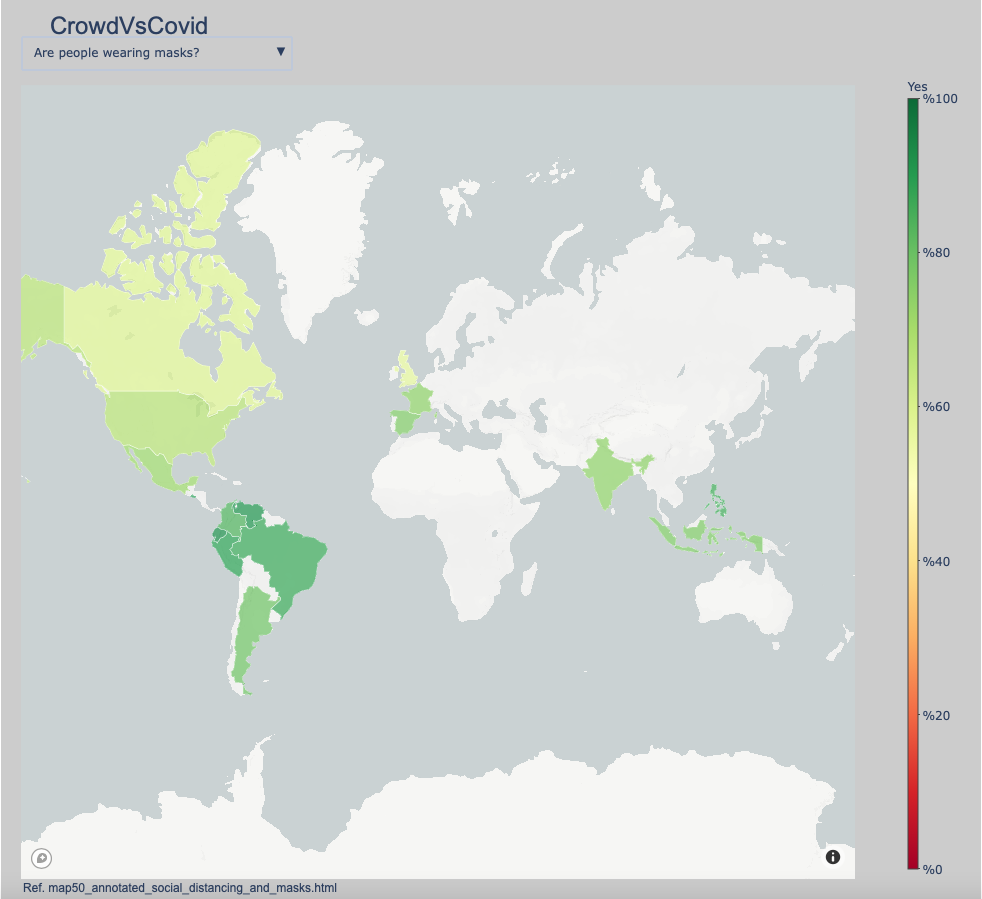}
\end{subfigure}
\caption{Image-based social-sensing results for social distance (left) and for face masks wearing (right) at mid May (top), early August (middle) and mid August, 2020 (bottom).}
\label{fig:summary}
\end{figure}

In Fig. \ref{fig:summary}, we show the maps produced for two different questions, i.e., \emph{``Are people respecting social distance?''} (left) and \emph{``Are people wearing face masks?''} (right) for the three time periods considered in the analysis so far (May 13, Aug. 1, and week 34 of 2020, i.e., August 17-23, 2020).

The maps are generated using the Python plotly library with Mapbox choropleth polygon maps and are interactive, allowing the user to select different questions for display and by hovering on a country to see the count statistics for each from which the percentage indicator has been computed.

\begin{table}[t]
\centering
\caption{Number of tweets after each phase of the social-sensing pipeline.}
\begin{tabular}{lr}
\textbf{stage of pipeline} & \textbf{\# tweets} \\
\hline
crawled tweets with images & 470,255 \\ 
after all image filtering    & 42,978 \\ 
after automated geolocating      & 25,541 \\ 
annotated via crowdsourcing   & 2,461 \\ 
with location validated by crowd       & 2,061 \\ 
\hline
\end{tabular}
\vskip 7pt
\label{tab:steps}
\end{table}

\section{Experiments}
\label{sec:datasets} \label{sec:experiments}

We now discuss an actual execution of the pipeline and analyse the results it produced.

A Twitter crawl we are going to analyze was performed on May 13, 2020 using the keywords: \{\emph{coronavirus, corona, virus, covid, covid19, covid-19, flu, wuhan, Coronaviridae, N95}\}.
The crawl was limited to tweets containing images and produced a total of 470,255 tweets, all posted within a 37-hour time period from May 12, 2020 02:02:06 to May 13, 2020 14:58:27 (GMT). After filtering the crawled images through the de-duplication + photo/NSFW/person/scene detection pipeline, a total of 42,978 tweets remained. Of those, only 3\% were natively geolocated, which is in line with percentages often reported in the literature. Using the CIME algorithm we were able to geolocate 25,541 tweets (59\%), which were used for the crowdsourcing phase.

In Table \ref{tab:steps} we show the number of tweets after each phase of the pipeline. We note that the number of tweets emerging from the geocoding step was too large to be evaluated in its entirety by the crowd available, and only around 10\% of the tasks were completed by at least three crowd workers. Possible remedies for this scaling issue will be discussed later, but we note that even with a limited crowd resource, we are able to produce reasonable predictions (as noted in Section~\ref{sec:external}) based on the completed tasks for which the crowd confirmed the geolocation of the image.



\subsection{Evaluating Individual Components}
\label{sec:individual}

We now discuss the validity of the approach. In this section we analyse the performance of each component of the pipeline (namely the image filters and the geocoding), and then in Section \ref{sec:external} we compare the obtained country-wise indicators with those derived from an external survey-based data source.

\subsubsection{Filter Evaluation}
Each filter described in Section \ref{filtering_classifiers} was evaluated by computing its Precision, Recall and $F_1$ measure as shown in Table~\ref{tab:filter_evaluation}. Precision in this case measures the percentage of relevant images among those retrieved (accepted) by the filter, while Recall is the fraction of relevant images retrieved.
Metrics were computed on a random sample of 700 images from the crawl, with ground truth annotations provided by three independent annotators and aggregated using majority-vote.
Test images for the NSFW, people, and photo detectors were selected from the entire crawl, while for the scene classifier, they were chosen from those filtered by the photo detector, since the scene detector expects to see only photos (rather than cartoons, etc.).

\begin{table}[t]
\centering
\caption{Evaluation of the various image filters in pipeline. Each filter is evaluated on 700 randomly chosen images with the ground truth labelled by 3 independent annotators.}
\begin{tabular}{lrrr}
\textbf{Filter} & \textbf{Precision} & \textbf{Recall} & \textbf{F1} \\ 
\hline
photo detector & 99.77\% & 94.67\% & 97.15\% \\
people detector & 95.81\% & 98.77\% & 97.26\% \\ 
NSFW detector & 99.38\% & 99.85\% & 99.61\% \\ 
public-private scene & 91.81\% & 96.91\% & 94.29\% \\ 
\hline
\end{tabular}
\vskip 1em
\label{tab:filter_evaluation}
\end{table}

\subsubsection{Geocoding evaluation}

One of the questions in the crowdsourcing task was evaluating the correctness of the geolocation. We excluded the posts locations which the crowd marked as ``surely wrong''.
From the data reported in Table \ref{tab:steps}, it can be seen that 84\% of the posts with automatically extracted locations could be retained for further analysis.

\subsection{External Validation}
\label{sec:external}

We  evaluated the accuracy of the overall pipeline by comparing the statistics generated from the annotated images with an external online-survey dataset. In particular, we compare with the CovidDataHub survey dataset \cite{coviddatahub}, since it provides  COVID-19 related social impact information on a weekly basis. 
The  aspect on which we focus for the validation is the use of masks in different countries, with the specific survey question being whether the respondent has \emph{``Worn a face mask outside my home''}, with possible answers: \emph{Not at all, rarely, sometimes, frequently, always}.
For the   considered periods the CovidDataHub portal provides data about a variable number of countries  (out of 29 in which surveys are run). 
The corresponding question in VisualCit is: \emph{``Are the people wearing masks?''} with reference to an image of people in a public location, where the possible answers were: \emph{Yes, Some of them, No, Cannot tell}. Images assigned the `Cannot tell' response were excluded from further analysis; to enable the comparison with the survey data, we map the survey responses down to three categories as follows: \emph{\{Not at all, rarely\} 	$\Rightarrow$ No, \{Sometimes\} 	$\Rightarrow$ Sometimes, \{Frequently, Always\} 	$\Rightarrow$ Yes}.

\begin{figure}[h]
\centering
  \includegraphics[width=0.8\columnwidth]{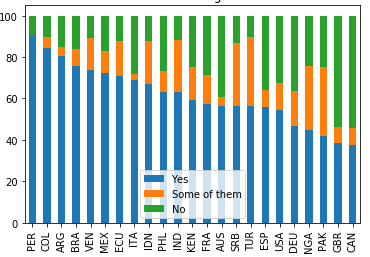}
  \label{fig:polimi4}
\caption{Are people wearing masks? Results from  VisualCit social media based pipeline. Here we include only countries with at least 50 labelled images from the crowd as of mid May, 2020.}
\label{fig:comparison4-2}
\end{figure}

For our image-based pipeline, we include only countries and territories for which we have collected at least 50 annotated images\footnote{The datasets used for the analysis are available: DOI 10.5281/zenodo.4539697}. This corresponded to 23 countries in total as shown in Figures~\ref{fig:summary} and \ref{fig:comparison4-2}. It should to be noted that the map includes a number of countries for which performing online surveys might be complicated for various reasons. 

\begin{figure}[h]
\centering
  \centering
  \includegraphics[width=0.8\columnwidth]{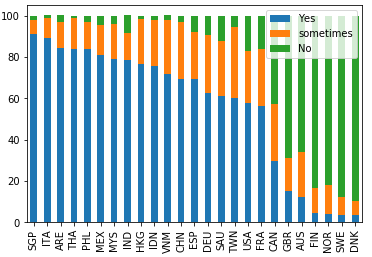}

\caption{Are people wearing masks? Results from CovidDataHub  surveys.}
\label{fig:comparison4-1}
\end{figure}

In Figures \ref{fig:comparison4-1} and \ref{fig:comparison4-2}, we compare charts of the survey results (CovidDataHub data) and the VisualCit pipeline for countries with more than 50 annotated images each. The visual correspondence between the two charts, which are based on completely independent data sources, lends a great deal of support to our approach.  
To further investigate the correspondence, we computed Pearson's correlation between the survey responses and the results of our social sensing pipeline for the countries in common in the two studies, as shown in Table~\ref{tab:correlations}. 

\begin{table}[t]
\caption{Correlation between Image-based social-media based frequency estimates  and survey responses  for Yes and No answers to the  question about face mask wearing (threshold 50 posts) and number of countries represented.}
\centering
\begin{tabular}{lcccccc}
\textbf{\shortstack{Crawl\\Date} }
& \textbf{\shortstack{Tweets\\number}}
& \textbf{\shortstack{\emph{Yes} \\ corr.}} 
& \textbf{\shortstack{\emph{No} \\ corr.}} & \textbf{\shortstack{Visual\\Cit}} & \textbf{\shortstack{Covid\\Data\\Hub}}
& \textbf{\shortstack{Com-\\mon}
}\\ 
\hline
May-13 &	3,605 &	0.91 &	0.85 &	10 &	25 & 7	\\
Aug-01	& 5,412 &\multicolumn{2}{c}{none in common} 	&		10	& 4	& 0\\
Aug 17-23& 5,944 &	0.18 &	0.63 &	15 &	29 & 9	\\
\hline
\end{tabular}
\vskip 1em
\label{tab:correlations}
\end{table}

\begin{figure}
  \centering
  \includegraphics[width=0.7\columnwidth]{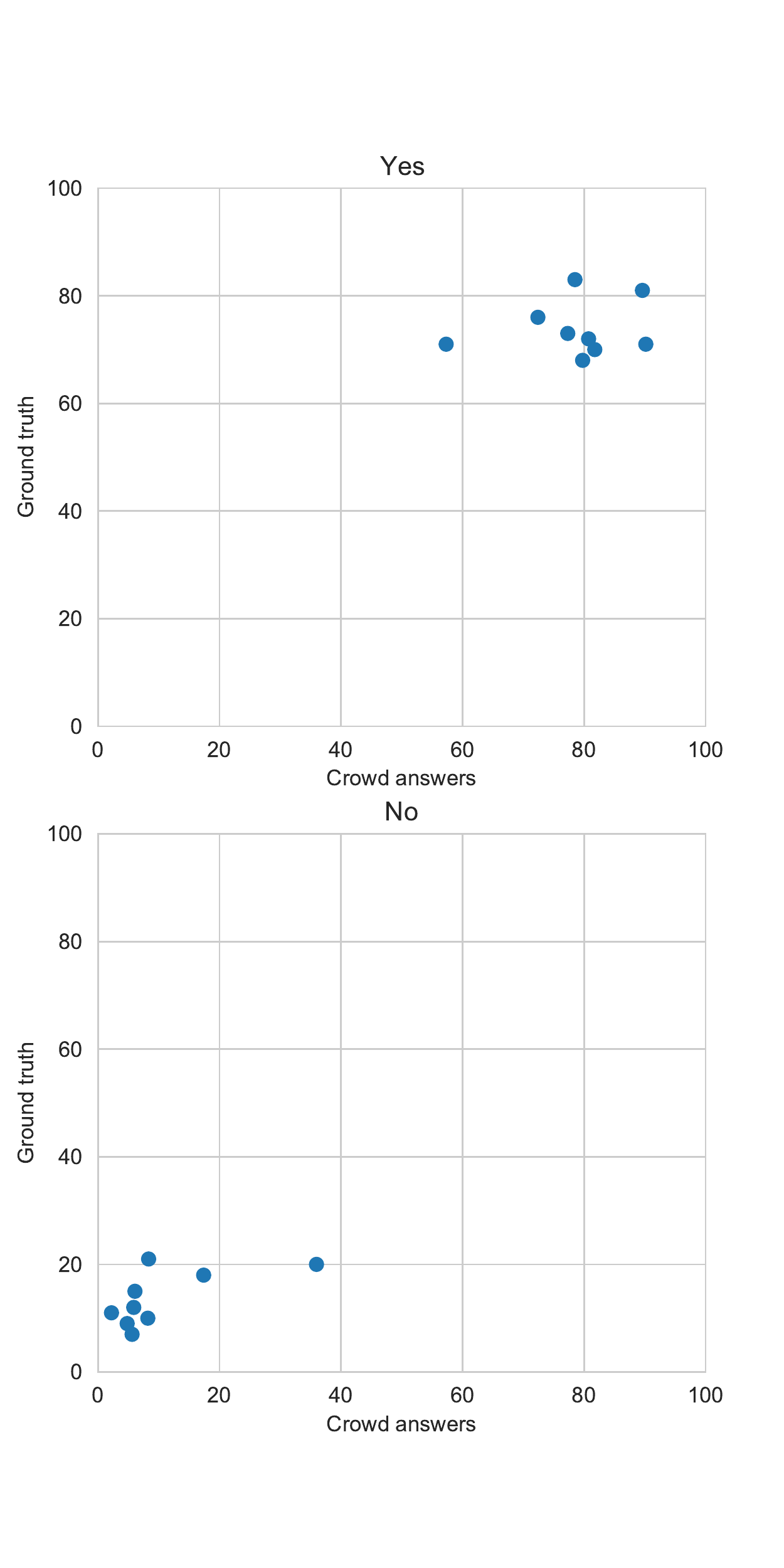}
 
\caption{Common countries evaluations in CovidDataHub as ground truth and image-based social sensing for May 17-23, with threshold set at 50}
\label{fig:correlation-plots}
\end{figure}

For the data crawled at the beginning of August a comparison could not be made since the two studies do not have any common country.
For Week 34 (Aug 17-23), while the general correlation in the Yes answer in Table  \ref{tab:correlations} is not present, in Figure \ref{fig:correlation-plots} we see how all the common countries in the surveys and in the image-based social sensing are evaluated, showing that most countries present similar values for the three classification of mask use, so the weak correlation is due to an outlier.

In conclusion, even with a limited number of annotations from the crowd we see that the indicators produced are well correlated with external survey data. It should be noted that the information gathered through surveys can themselves be variable. For instance, for the week July 27-August 2, 2020, the CovidDataHub shows data only for four countries out of the 30 initially considered in the survey, With our method the number of covered countries should be less dependent on the period and crowd workers can help providing some useful indications to decision makers (see Fig. \ref{fig:summary}, showing all countries with the threshold set at 50 posts).

\subsection{Limitations of the proposed approach}

The proposed pipeline faces the following limitations:  (1) crowd availability and cost to classify the social media images based on people behaviour; (2) computational constraints due to the geolocation algorithm (CIME) and the execution of the different computational models (Filters).


Availability of a committed crowd within the given time \cite{kirilenko2017crowdsourcing} and vast volumes of data \cite{rudra2016summarizing} is often a challenge and a limiting factor for crowdsourcing initiatives. While federating existing communities \cite{ravi2019crowd4ems} can ensure commitment, the volume of tasks could hinder the community in sustaining the interest. Volunteer crowdsourcing initiatives are not incentivized and are often motivated by factors including, but not limited to altruism, peer-indulgence, curiosity, fun and in some cases because of the organization that conducts the initiative. Usually, the lack of motivation can be a result of tasks being either highly difficult or mundane. 
Paid crowdsourcing initiatives are also a common practice in outsourcing micro-tasks. Paid crowdsourcing platforms like Amazon Mechanical Turk are designed to handle equally complex tasks and provide a timely, scalable workforce. Studies reveal that higher redundancy and limiting the focus to a specific geographic area can achieve a similar quality of result output \cite{borromeo2016investigation,kirilenko2017crowdsourcing} in a much lesser time when compared to the volunteering crowd. Future work will address the quality control mechanisms in designing the experiment in a paid crowdsourcing platform.   

The ability to analyze the tweets with VisualCit could also be limited in the first image filtering and geolocation steps by the amount of computational resources available when the number of tweets to be analyzed is high. Some improvements can be gained using more efficient algorithms if available or training new specific classifiers.


\section{Concluding remarks}
\label{sec:conclusions}

In this paper we have demonstrated that it is possible to build a social sensing pipeline with humans in-the-loop for collecting important policy indicators from social media images posted on Twitter. 
The presented approach is general and can be extended to other contexts of investigation, selecting the appropriate filters and questions to the crowd. 

In order to derive indicators, the main difference between conduction online surveys and crowdsourcing to annotate evidence collected from social media is in the number and role of persons to be involved. While for survey a large sample is needed, for image-base social sensing with crowdsourcing a much smaller number of persons can be involved. 

There are a number of directions for improving the image-based social sensing pipeline, including: 
\begin{itemize}
    \item Multi-linguality: extend the crawl to languages such as Arabic, Russian, Spanish, Portuguese, Indonesian, etc.
    \item Evaluate the approach for new indicators, such as monitoring climate events (such as floods, hurricanes, etc.) or engagement in social movements (such as the "Black Lives Matter" protest movement).
    \item Study feedback mechanisms and non-monetary incentives to increase crowd involvement and overcome scaling issues.
    \item Make use of the crowd responses to fine-tune specific filters and improve their performance over time.
    \item Incorporate more sophisticated crowdsourcing quality control mechanisms, such as worker vetting.  
    \item Develop methods to evaluate the confidence of the obtained results in a systematic way.
\end{itemize}

This paper does not address the potential malicious usage of social media data analysis for society surveillance, nor privacy issues that could arise from the analysis publicly available social media data. These concerns will be addressed in future research.

\section*{Acknowledgments}
This work was partially funded by the European Commission H2020 project  Crowd4SDG  ``Citizen Science for Monitoring Climate Impacts and Achieving Climate Resilience'', project no. 872944. 
The authors thank all students working as crowd workers and all other anonymous contributors to the crowdsourcing phase.
The proof-of-concept prototype for this research was started during the VersusVirus and EUvsVirus online hackathons run in April and May 2020.
The authors thank Citizen Cyberlab (\url{https://citizencyberlab.org/}) for the organisation of the session at VersusVirus and EUvsVirus hackathons, and the Citizen Science Center (\url{https://citizenscience.ch/en/}) for providing the Pybossa software platform. 

\section*{Appendix A: Crowdsourcing questions} \label{sec:appendix}

{\small
Questions asked during the crowdsourcing step were the following, with some questions contingent on the answer to the previous question.  
\begin{enumerate}
    \item Is this a photo (rather than a cartoon, graph, meme, etc.)? \\
    - Yes, No, Not Sure
    \item Does it look like it has been taken recently (in the last three months)? \\
    - Yes, No, Cannot tell
    \item Are there people in this image?\\
    - Yes, No, Not Sure
    \item Are the people wearing masks? \\
    - Yes, Some of them, No, Cannot tell
    \item If so, which type? \\
    - Scarf, Cloth, Surgical, FP2, FP3, Gas mask, Other, Cannot tell
    \item Are the people wearing the mask correctly? \\
    - Yes, No, Only some of them, Cannot tell, Not sure
    \item How many people are there in the image? \\
    - 1, 2, 3, 4, 5 or more
    \item Are they respecting social distance? \\
    - Yes, No, Cannot tell.
    \item Are they in a public place (shops, outdoors, ...)? \\
    - Yes, No, Not sure
    \item If they are in a public place, what type? 
    - street/square, park, shop, hospital, outdoors, other, cannot tell
    \item What are the people doing? \\
    - socializing, exercizing, shopping, queuing, volunteering, protesting, working, other, cannot tell
    \item We have associated a country or territory with this image. Do you think the picture was likely taken in this location? \\
    - Yes, Maybe, Surely not, Cannot tell.
\end{enumerate}
}


\end{document}